\documentclass[pra,aps,twocolumn]{revtex4}
\usepackage{epsfig}

\usepackage{graphicx}

\begin{document}


\title{Ginzburg-Landau equation bound to the metal-dielectric interface
and transverse nonlinear optics with amplified plasmon polaritons}

\author{A. Marini and D.V. Skryabin$^*$}
\affiliation{Centre for Photonics and Photonic Materials,
Department of Physics, University of Bath, Bath BA2 7AY, UK\\
$^*$Corresponding author: d.v.skryabin@bath.ac.uk}
\date{\today}
\begin{abstract}
Using a multiple-scale  asymptotic approach,  we have derived the
complex cubic Ginzburg-Landau equation for  amplified and
nonlinearly saturated surface plasmon polaritons propagating  and
diffracting along a metal-dielectric interface.  An important feature of our method
is that it explicitly accounts for nonlinear terms in the boundary
conditions, which are critical for a correct description of 
nonlinear surface waves. Using our model we have analyzed
filamentation  and discussed  bright and dark spatially localized structures
of plasmons.
\end{abstract}

\maketitle


\section{Introduction}
Surface plasmon polaritons (SPPs) are half-photon half-electron
surface waves. Thanks to their dual nature, SPPs can be focused
tighter than pure light, which is an important property
for potential applications in optical processing of information. In the absence of lateral
boundaries propagating SPPs are expected to diffract in the
interface plane. One way to control diffraction is to structure the
surface and make plasmonic waveguides, see, e.g.,
\cite{Bozhevolnyi2006}. Alternatively, one can use nonlinearity for
the creation of spatial SPP solitons, which are non-diffracting self-localized surface waves
\cite{Feigenbaum2007,Davoyan2009}. Other transverse
effects with nonlinear plasmons, such
as self-focusing and filamentation \cite{Lin2009}
can be important  for frequency conversion, switching and routing
experiments with photonics chips. Note, that the
interplay between transverse and nonlinear effects have attracted
significant  attention outside the plasmonics and
nanophotonics contexts, see, e.g., \cite{ABRAHAM1990,Kivshar2003}
for the historic accounts.

Nonlinear functionality of SPPs
\cite{Wurtz2006,Pacifici2007,MacDonald2009} can be significantly
hampered by ohmic losses resulting in short propagation distances.
One of the possible solutions is to amplify SPPs by doping and
pumping the dielectric, so that the losses are either partially or
fully compensated \cite{Winter2006,Noginov2008,Ambati2008}. The
linear dispersion of the amplified SPPs  has been studied by several
groups, see, e.g., \cite{Nezhad2004,Noginov2008a}. The linear
results have been recently generalized to the more realistic case,
when linear gain is nonlinearly saturated above the stimulated
emission threshold \cite{Marini2009}.

Analytical or semi-analytical approaches  to describe nonlinear
effects with  SPPs are very important, since the
first principle numerical modelling of
nonlinear and multidimensional cases is still computationally
demanding. Recently the nonlinear Shrodinger equation (NLS) has been
introduced for the plasmons in a slot waveguide formed by two
planar metal dielectric interfaces
\cite{Feigenbaum2007} and at a single interface
\cite{Davoyan2009}. The {\em averaging method} implemented in
\cite{Feigenbaum2007,Davoyan2009} has been borrowed from the theory
of the dielectric waveguides \cite{Agrawal2001}. In this approach,
one starts from the known solution for the linear SPPs: $\vec
F(x)e^{i\beta z}$, where $x$ is the coordinate perpendicular to the
interface, $z$ is the propagation direction and $\beta$ is the
propagation constant. Then introducing a slowly varying amplitude
$A(z,y)$ and assuming small nonlinearity, the  Maxwell
equations are averaged in $x$ and the  NLS equation for
$A$ is derived \cite{Feigenbaum2007,Davoyan2009}.

The above approach has some drawbacks. First, it is sufficiently
well justified only for the quasi-transverse fields, that approximately
satisfy the wave equation \cite{Agrawal2001}. Another problem is
that it does not  treat the boundary conditions rigorously.
In particular,  continuity of the normal to the
interface component of the displacement $D_x$ is guaranteed only in the linear
approximation.
If the intensity of the guided light  peaks away from the
interfaces and is small in the proximity of the latter (like it
typically happens in dielectric waveguides operating on the
principle of total internal reflection), then the nonlinear terms in
the boundary conditions can  be disregarded. However, the SPP
intensity peaks exactly at the interface and nonlinear contribution to the 
boundary conditions is critical \cite{Marini2009,Mihalache1987}. This
is also true for other types of nonlinear surface waves, see,
e.g., \cite{Agranovich1980,BOARDMAN1986,Moloney1992}. 

Thus, it is important to
develop a  rigorous procedure of deriving the nonlinear
evolution equation for  surface plasmons,
such that the continuity of $D_x$ is enforced
together with its nonlinear part. 
Below we  develop a multiple-scale asymptotic  approach for 
the amplified SPPs, which treats the nonlinear boundary conditions rigorously and reveals the
differences with the results obtained by the averaging technique.
The gain and complex nonlinearity we use are derived using the two-level model.
Our procedure leads to the complex
Ginzburg-Landau equation that accounts  for diffraction of SPPs in the interface
plane. The nonlinearity enhancement factor derived by us is
intrinsically complex, whilst the averaging approach gives a real one.
The difference between the predictions of two approaches increases
in the short wavelength limit, where  SPPs are maximally localized
at the interface, and thus the nonlinear part of the boundary
condition is more important. Using our theory we derive a
criterion for  SPP filamentation and  discuss bright  
and dark  spatially localized  SPPs. Strictly speaking 
both soliton families are unstable, though the bright solitons demonstrate propagation distances
sufficient for their practical observation.

\section{Model}
We assume that the interface between metal and dielectric is
at $x=0$ and $z,y$ are  the in-plane coordinates.
The evolution of  SPPs  obeys
the time independent Maxwell equations
\begin{eqnarray}
\label{max1} && \partial^2_{xy}E_y-\partial^2_{yy}
E_x-\partial^2_{zz} E_x
+ \partial^2_{zx} E_z=\epsilon  E_x,\\
\label{max2} &&
\partial^2_{yz} E_z-\partial^2_{zz} E_y-\partial^2_{xx} E_y
+ \partial^2_{xy} E_x=\epsilon  E_y,\\
\label{max3} && \partial^2_{xz} E_x-\partial^2_{xx}
E_z-\partial^2_{yy} E_z + \partial^2_{zy} E_y=\epsilon  E_z.
\end{eqnarray}
The coordinates are dimensionless and normalized to the inverse wavenumber
$k=2\pi/\lambda_{vac}$, where $\lambda_{vac}$ is the vacuum
wavelength. The permittivity on the dielectric side of the interface ($x>0$) is
\begin{eqnarray}
&& \epsilon=\epsilon_d+\chi (|E_x|^2+|E_y|^2+|E_z|^2),\\
&& \epsilon_d=\epsilon^{\prime}_d+i\epsilon_d^{\prime\prime},~~
\chi=\chi'+i\chi''.
\end{eqnarray}
The permittivity on the metal side ($x<0$) is
\begin{equation}
\epsilon=\epsilon_m=\epsilon^{\prime}_m+i\epsilon_m^{\prime\prime}.
\end{equation}

If SPPs are amplified by means of  active inclusions in the dielectric, 
then  $\epsilon_d$ and $\chi$ are
functions of the gain coefficient $\alpha$. The propagation constant
$\beta$ for linear plasmons is
\begin{equation}
\beta=\sqrt{\epsilon_d\epsilon_m /(\epsilon_d+\epsilon_m)}.\label{bet}
\end{equation}
where $\beta$ becomes real at the threshold: $\alpha=\alpha_0$  \cite{Nezhad2004}:
\begin{equation}
\beta(\alpha_0)\equiv\beta_0, ~~Im\beta_0=0.
\end{equation}
The linear and nonlinear permittivities for the dielectric  at $\alpha=\alpha_0$ are
$\epsilon_d(\alpha_0)=\epsilon_{d0}$ and $\chi(\alpha_0)=\chi_0$.

The active inclusions are approximated by the 2-level atom susceptibility. For
light intensities  much smaller than the transition saturation
intensity $I_s$ we find \cite{Marini2009,Boyd2003}
\begin{eqnarray}
\epsilon_d & = & \epsilon_b-\alpha\frac{i-\delta}{1+\delta^2}, \\
\chi & = & \alpha\frac{i-\delta}{(1+\delta^2)^2}.
\end{eqnarray}
where $\epsilon_b$ is the real dielectric constant of the background
material hosting the two-level atoms. $\delta=(\omega-\omega_a)T_2$
is the dimensionless detuning from the atomic resonance frequency,
$\omega_a=2\pi c/\lambda_a$, normalized to the transition linewidth,
$T_2^{-1}$. $\alpha$ is the dimensionless gain coefficient at the line
center. The electric field is normalized to
$\sqrt{I_s}$, which implies that the
nonlinear susceptibility $\chi$ is dimensionless, see
\cite{Marini2009} for more details. Possible dependence of the atomic life times
from the distance to the interface, see, e.g., \cite{Leon2008}, are specific to 
a choice of a pumping technique and are disregarded in what follows.

\begin{figure}
\centerline{\includegraphics[width=0.45\textwidth]{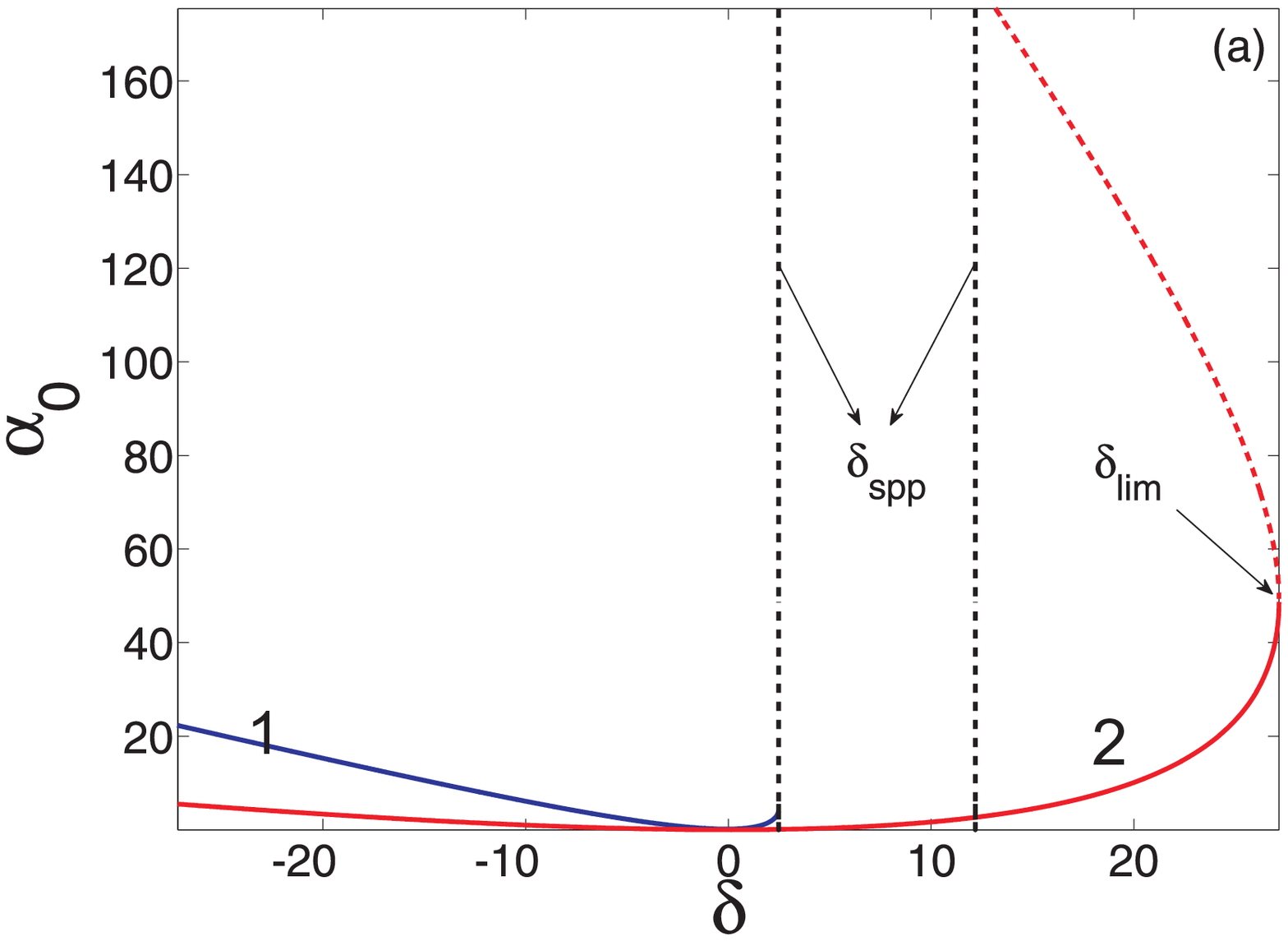}}
\centerline{\includegraphics[width=0.45\textwidth]{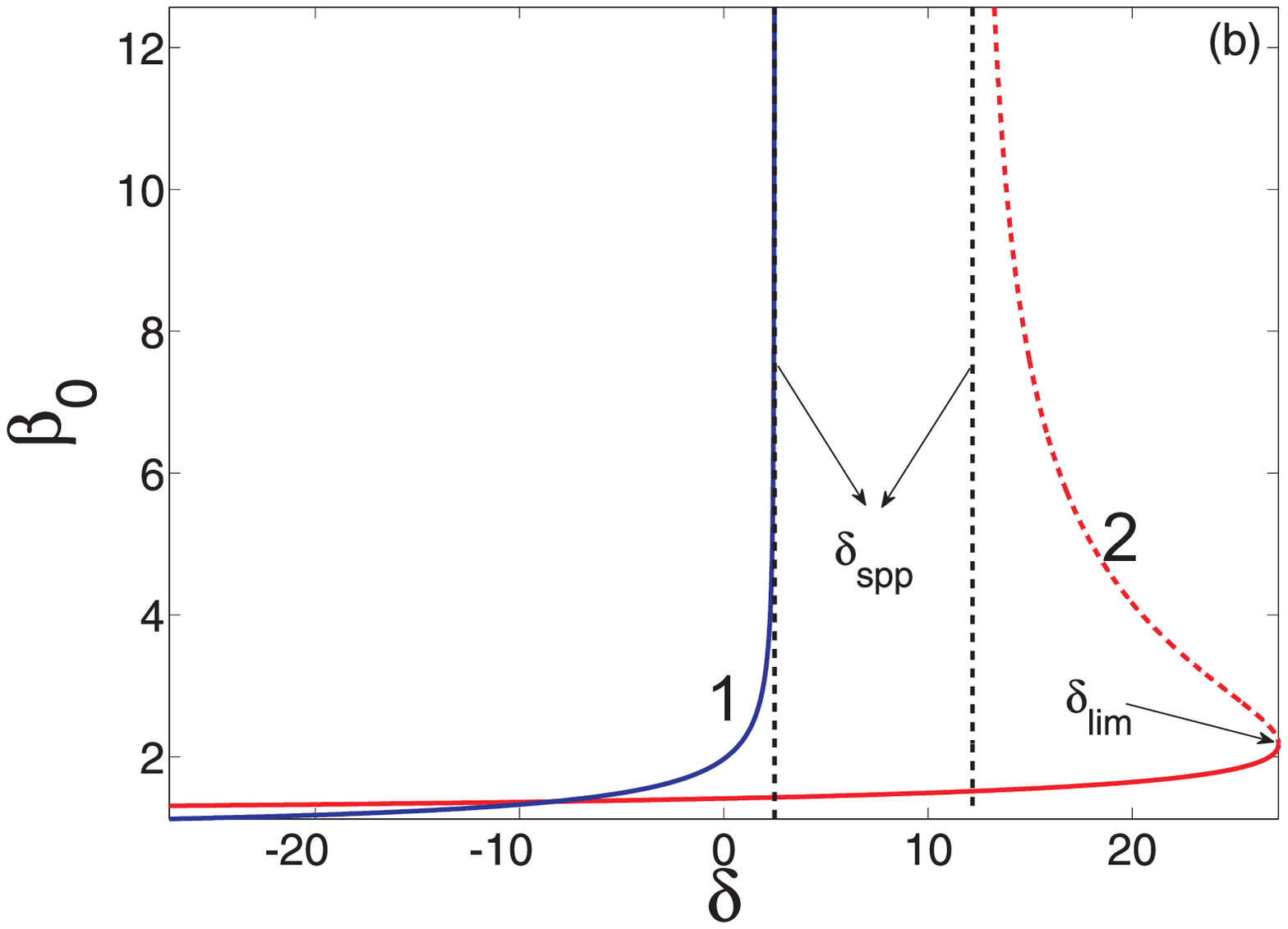}}
\caption{(Color online) (a) Threshold gain $\alpha_0$ vs detuning $\delta$ for two
different atomic resonances:  $\lambda_a=400$nm (line 1 (blue)) and
$\lambda_a=700$nm (line 2 (red)). (b) $\beta_0$ vs $\delta$. Parameters
and notations as in a). The full/dashed line corresponds to the
minus/plus in Eq. (\ref{LinG}).} \label{alpha}
\end{figure}

The threshold gain $\alpha_0$ works out as
\begin{eqnarray}
\nonumber && \alpha_0(\omega) = \frac{1}{2\epsilon''_m}\left(\left|\epsilon_m\right|^2-2\epsilon''_m\epsilon_b\delta\right)\\
 && \pm \frac{1}{2\epsilon''_m}\sqrt{\left|\epsilon_m\right|^4-4\epsilon''_m\epsilon_b\left(\epsilon''_m\epsilon_b+\delta\left|\epsilon_m\right|^2\right)}.
\label{LinG}
\end{eqnarray}
Lossless propagation of SPPs is impossible above the critical value
of $\delta=\delta_{lim}$ as determined by the condition that the square
root in Eq. (\ref{LinG}) becomes zero. At this point, the two
solutions for $\alpha_0$   degenerate, see Fig. \ref{alpha}.
$\delta_{lim}$ should not be confused with the plasmon resonance
frequency, $\delta_{spp}$, which corresponds  to the  zero of the
denominator of $\beta_0$. The  existence boundaries  of the SPPs
at $\alpha=\alpha_0$ are  determined by either or both of $\delta_{lim}$ and
$\delta_{spp}$, see Fig. \ref{alpha}. The upper branch solution
(dashed lines in Fig. \ref{alpha}) corresponds to   high gain
coefficients implying refractive indices of  order 10 or larger.
In our subsequent numerical examples we focus on  relatively
small $|\delta|$'s, thereby selecting the  lower branch of $\alpha_0$
(minus sign in front of the square root in Eq. (\ref{LinG})). This branch corresponds
to relatively small changes of the background refractive index
as achievable for small densities of active atoms \cite{Noginov2008a}.
Note, that $\epsilon_m$ is frequency dependent, i.e., $\epsilon_{m}(\omega)=\epsilon_m(\delta/T_2+\omega_a)$.
Hence $\alpha_0$ is a function of both $\delta$ and $\omega_a$.
We choose silver as a metal in all our calculations.

\section{First principle derivation of the Ginzburg-Landau equation for SPPs}
The perturbation theory developed below assumes relatively
small deviations of the gain coefficient from its threshold value $\alpha_0$,
i.e.
\begin{equation}
\alpha-\alpha_0\equiv \alpha_0 g, ~~|g|\ll 1.
\end{equation}
The anzats for the field components in the dielectric and metal is
\begin{eqnarray}
\nonumber && E_{x,j}=\left[A_{x,j}^{(0)}+A_{x,j}^{(1)}+O(|g|^{5/2})\right]e^{i\beta_0z},\\
\label{sub} && E_{y,j}=\left[A_{y,j}^{(0)}+O(|g|^{2})\right]e^{i\beta_0z},\\
\nonumber && E_{z,j}=\left[A_{z,j}^{(0)}+A_{z,j}^{(1)}+O(|g|^{5/2})\right]e^{i\beta_0z}, ~j=d,m
\end{eqnarray}
where $A_{x,j}^{(0)}\sim |g|^{1/2}$, $A_{x,j}^{(1)}\sim |g|^{3/2}$, $A_{y,j}^{(0)}\sim |g|$,
$A_{z,j}^{(0)}\sim |g|^{1/2}$, and $A_{z,j}^{(1)}\sim |g|^{3/2}$.
All   $A$'s are the functions of $z,y$ and $x$. However,
their dependencies  on $z$ and $y$ are assumed slow relative to the fast oscillations of
$e^{i\beta_0z}$:
\begin{equation}
\partial_y\sim |g|^{1/2},~\partial_z\sim |g|.
\end{equation}
Though the $y$ component  deviates from zero, when  the field has finite
size along $y$, it is expected to remain relatively small, $\sim |g|$.

The dielectric susceptibility is trivially expanded into the $g$-series:
\begin{eqnarray}
&& \epsilon_d=\epsilon_{d0}+\epsilon_{d1},~\epsilon_{d1}\equiv g\alpha_0\partial_{\alpha}\epsilon_d.
\end{eqnarray}
The only value of the nonlinear coefficient $\chi$, we require below is the one
taken exactly at the threshold $\chi(\alpha_0)\equiv\chi_0$.

\subsection{$|g|^{1/2}$ and $|g|^{1}$ orders}
By substituting Eqs. (\ref{sub}) into the Maxwell equations, we find in the  order $|g|^{1/2}$
\begin{eqnarray}
\label{m0}
\hat M_j\left[\begin{array}{c}  A_{x,j}^{(0)} \\  A_{z,j}^{(0)} \end{array}\right]=0,~j=d,m.
\end{eqnarray}
Here
\begin{eqnarray}
\label{m1}
\hat M_j=\left(\begin{array}{cc}  q_j^2 & i\beta_0\partial_x \\ 0 &
\partial^2_{xx}-q_j^2 \end{array}\right)
\end{eqnarray}
and
\begin{equation}
q_d^2=\beta_0^2-\epsilon_{d0},~q_m^2=\beta_0^2-\epsilon_{m}.
\end{equation}
Any nonlinear and transverse, i.e., $y$-dependent, effects are disregarded in
Eq. (\ref{m0}).

The SPP solution of Eq. (\ref{m0}) is well known:
\begin{eqnarray}
\nonumber A_{x,d}^{(0)} & = & \frac{i\beta_0}{q_d}A(y,z)e^{-q_dx}, \\
\label{e0d} A_{z,d}^{(0)} & = & A(y,z)e^{-q_dx}, \\
\nonumber A_{x,m}^{(0)} & = & -\frac{i\beta_0}{q_m}A(y,z)e^{q_mx}, \\
\nonumber A_{z,m}^{(0)} & = & A(y,z)e^{q_mx}.
\end{eqnarray}
Eqs. (\ref{e0d}) satisfy continuity of the normal component of the displacement and  tangential
components of the field:
$\epsilon_d A^{(0)}_{x,d}=\epsilon_m A^{(0)}_{x,m}$
and $A^{(0)}_{z,m}=A^{(0)}_{z,d}$ at $x=0$. The former condition implies
$\epsilon_{d0}q_m=-\epsilon_mq_d$ giving (after some algebra)
the expression for $\beta_0$, see Eq. (\ref{bet}).

In the order $|g|^{1}$, we find the linear equations for the $y$ component of the field
\begin{equation}
\label{z2}  q_j^2A_{y,j}^{(0)}-\partial_{xx}^2A_{y,j}^{(0)}=0,
\end{equation}
which are readily solved
\begin{eqnarray}
\label{e0m} A_{y,d}^{(0)} & = & B(y,z)e^{-q_dx} \\
\nonumber A_{y,m}^{(0)} & = & B(y,z)e^{q_mx}.
\end{eqnarray}
To determine the unknown functions $A(y,z)$ ($|A|\sim |g|^{1/2}$) and $B(y,z)$ ($|B|\sim |g|$)
we need to proceed to the higher orders of our perturbation series.

\subsection{$|g|^{3/2}$ order and Ginzburg-Landau equation}
Proceeding to the order $|g|^{3/2}$, we find  an inhomogeneous system
of differential equations for corrections to the standard SPP solutions.
The correction equations on the metal side are
\begin{eqnarray}
\label{met} \hat M_m\left[\begin{array}{c}  A_{x,m}^{(1)} \\  A_{z,m}^{(1)} \end{array}\right]=
\left[\begin{array}{c}  K_x \\  K_z \end{array}\right]e^{q_m x},
\end{eqnarray}
where \begin{eqnarray}
&&K_x=\frac{\beta_0^2+\epsilon_m}{q_m}\partial_z A
-q_m\partial_yB-\frac{i\beta_0}{q_m}\partial_{yy}^2A,\nonumber\\
\nonumber &&K_z=-2i\beta_0\partial_zA-\partial_{yy}^2A.
\end{eqnarray}
A solution of Eqs. (\ref{met}) consists of a particular solution of
the  inhomogeneous problem plus a general solution of the
corresponding homogeneous system ($K_{x,z}=0$):
\begin{eqnarray}
&&A_{x,m}^{(1)}=\frac{1}{2q_m^3}\left[-i\beta_0K_z(1+q_mx)+2q_mK_x\right]e^{q_mx}\nonumber\\
\nonumber && -c\frac{i\beta_0}{q_m}e^{q_mx}\\
\label{e1m} &&A_{z,m}^{(1)}=\frac{K_z}{2q_m}xe^{q_mx}+ce^{q_mx},
\end{eqnarray}
where $c$ is a constant to be determined from the boundary conditions.

The righthand sides of the corresponding equations in the dielectric are more cumbersome due to
nonlinear terms:
\begin{eqnarray}
\label{die} && \hat M_d\left[\begin{array}{c}  A_{x,d}^{(1)} \\  A_{z,d}^{(1)} \end{array}\right]=\\ &&
e^{-q_d x}\left\{\left[\begin{array}{c}  L_x \\  L_z \end{array}\right]
+ \left[\begin{array}{c}  N_x \\  N_z \end{array}\right] e^{-2x\mathrm{Re}q_d}\right\},\nonumber
\end{eqnarray}
where
\begin{eqnarray}
\nonumber && L_x=-\frac{1}{q_d}(\beta_0^2+\epsilon_{d0})\partial_zA +\frac{i\beta_0}{q_d}\epsilon_{d1}A+    \\
\nonumber &&+q_d\partial_yB+\frac{i\beta_0}{q_d}\partial_{yy}^2A  \\
\nonumber && L_z=-2i\beta_0\partial_zA-\epsilon_{d1}A-\partial_{yy}^2A   \\
\nonumber && N_x=\frac{i\beta_0}{q_d}\left(\frac{\beta_0^2}{|q_d|^2}+1\right)\chi_0|A|^2A       \\
\nonumber && N_z=-\frac{ \left(\epsilon_{d0}q_d + 2\beta_0^2\mathrm{Re}q_d \right)\left( |q_d|^2+\beta_0^2 \right)}
{\epsilon_{d0}q_d|q_d|^2}\chi_0|A|^2A.
\end{eqnarray}
Solutions of Eqs. (\ref{die}) are given by
\begin{eqnarray}
&& A_{x,d}^{(1)}=\frac{1}{2q_d^3}\left[i\beta_0L_z(1-q_dx)+2q_dL_x\right]e^{-q_dx}\label{e1d} \\
&& +\frac{1}{q_d^2}\left[N_x+\frac{i\beta_0N_z(2\mathrm{Re}q_d+q_d)}{4\mathrm{Re}q_d(\mathrm{Re}q_d+q_d)}\right]
e^{-2\mathrm{Re}q_dx-q_dx} \nonumber \\
&& A_{z,d}^{(1)}=\left[-\frac{L_z}{2q_d}x
+\frac{N_ze^{-2\mathrm{Re}q_dx}}{4\mathrm{Re}q_d(\mathrm{Re}q_d+q_d)}\right]e^{-q_dx}. \nonumber
\end{eqnarray}
The arbitrary constant terms in Eqs. (\ref{e1d}) have been omitted,
as this does not lead to any loss of generality.

Combining Eqs. (\ref{e0d}),(\ref{e0m}),(\ref{e1m}),(\ref{e1d}) with  Eqs. (\ref{sub})
and substituting the calculated fields into the boundary conditions
\begin{eqnarray}
&& \nonumber [\epsilon_d+\chi (|E_{x,d}|^2+|E_{y,d}|^2+|E_{z,d}|^2)] E_{x,d}\\
&& =\epsilon_m E_{x,m},\label{bcond} \\ &&\nonumber~E_{z,m}=E_{z,d},~ E_{y,m}=E_{y,d},
\end{eqnarray}
we find that the latter are satisfied in the
order $|g|^{3/2}$ only providing that
\begin{equation}
c=\frac{N_z}{4(q_d+\mathrm{Re} q_d)\mathrm{Re} q_d} \label{W}
\end{equation}
and the  amplitude $A$ solves the  complex Ginzburg-Landau equation
\begin{eqnarray}
&& 2i\beta_0\partial_zA+\partial^2_{yy}A+fA+\gamma|A|^2A=0, \label{gl}
\end{eqnarray}
where
\begin{eqnarray}
\nonumber && f\equiv g\frac{\alpha_0\epsilon_m^2 \partial_{\alpha}\epsilon_{d}}{(\epsilon_{d0}+\epsilon_m)^2},\\
\nonumber && \gamma\equiv h\chi_0,~h\equiv\frac{\beta_0^4}{\epsilon_{d0}^2}
\frac{q_d(|q_d|^2+\beta_0^2)}{(q_d+\mathrm{Re} q_d)\left|q_d\right|^2}.
\end{eqnarray}
All the terms containing  $B(y,z)$ cancel out leaving this function undetermined
until the higher order corrections are accounted for. Thus taking the plasmonic
field as in Eqs. (\ref{sub})  with the amplitude $A$ obeying Eq. (\ref{gl}) we are guaranteed that the
nonlinear boundary conditions are satisfied upto and including the  $|g|^{3/2}$-terms.

The first and second terms in Eq.(\ref{gl}) describe the propagation and
diffraction of SPPs. ${\mathrm Re} f$ accounts for the shift of the propagation constant
away from $\beta_0$, when  gain deviates from the threshold. ${\mathrm Im} f$
accounts for the gain excess ($\alpha>\alpha_0, ~{\mathrm Im} f<0$)
or shortage ($\alpha<\alpha_0, ~{\mathrm Im} f>0$).
The nonlinear term provides an additional shift of the propagation constant
($\mathrm{Re}\gamma |A|^2$) and of the nonlinear loss ($\mathrm{Im}\gamma |A|^2$)
that counterbalances the linear gain. Note, that the transformation back to physical units
results in the appearance of a $k^2$ factor in the 3rd and 4th terms of Eq. (\ref{gl}).

\subsection{Comparison with the averaging approach}

Parameter $h$ in the expression for $\gamma$
is the nonlinearity enhancement factor. $h$ accounts for the
difference between the nonlinear responses of SPPs and free waves
propagating far from the interface.  $h$ is complex, therefore, even
when atomic nonlinearity is purely dispersive or purely absorptive,
the effective SPP nonlinearity is a mixture of both types. This contrasts
results using  the averaging approach \cite{Feigenbaum2007,Davoyan2009,Agrawal2001},
where $h$ is  real.

The averaging approach  yields a well known  expression
for the  effective nonlinearity of  guided modes.
Following this method, one should replace $h$ in Eq. (\ref{gl})
with $\tilde h$ \cite{Feigenbaum2007,Davoyan2009}
\begin{equation}
\tilde h=\frac{\int_{0}^{+\infty}dx\left|\vec{F}\right|^4}{\int_{-\infty}^{+\infty}dx\left|\vec{F}\right|^2}
\label{aver_h}\end{equation}
Here $\vec F$ is the plasmonic field given by Eq. (\ref{e0d}) with $A=1$.

\begin{figure}
\centerline{
\includegraphics[width=0.45\textwidth]{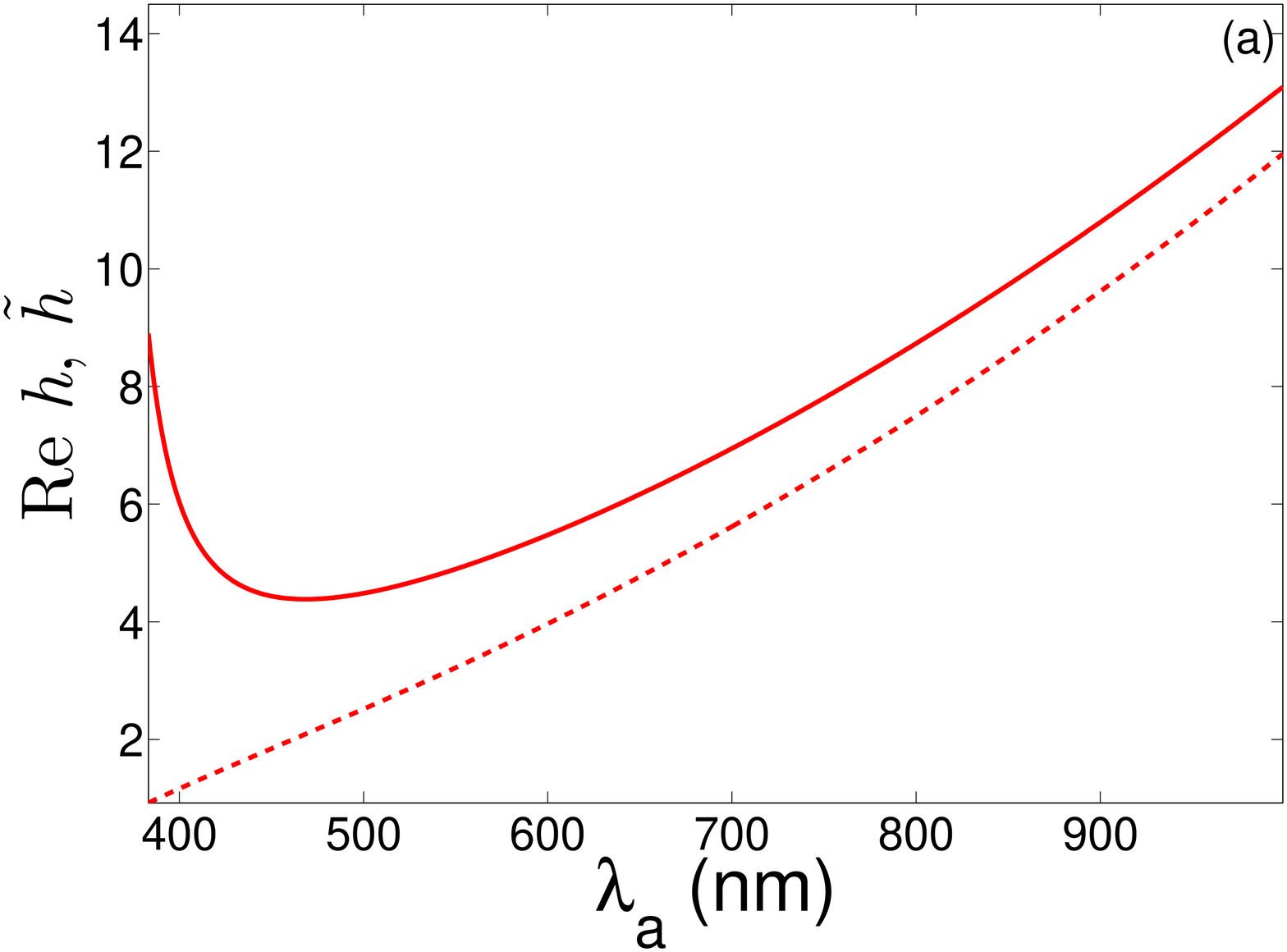}}
\centerline{\includegraphics[width=0.45\textwidth]{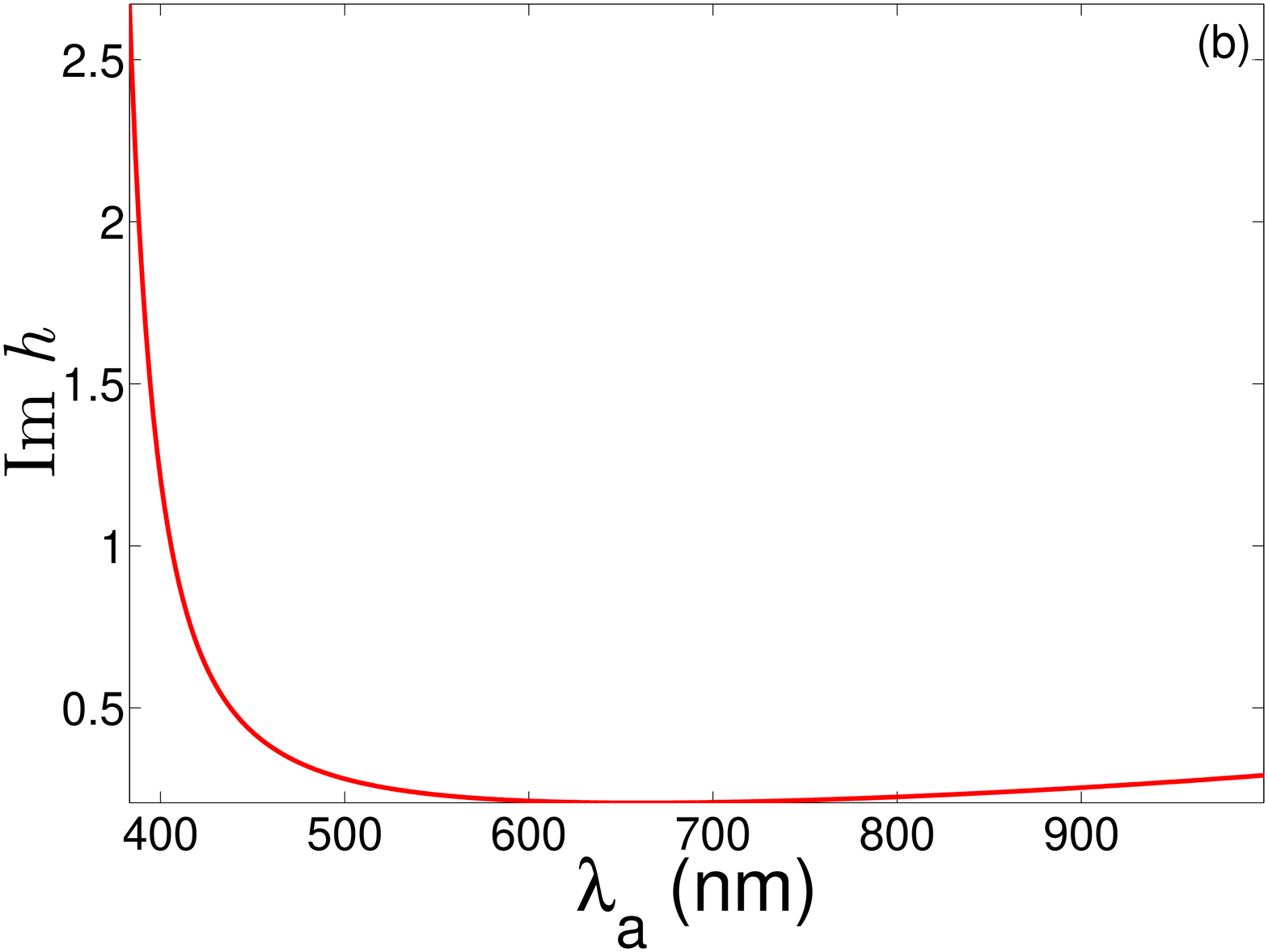}
}
\caption{(Color online) 
(a) Nonlinearity enhancement coefficients ${\mathrm Re} h$ (full line) and $\tilde h$ (dashed line)
calculated using two different approaches vs $\lambda_a$. (b) ${\mathrm Im}h$.
Varying $\delta$ inside the transition linewidth leads only to small difference in $h$.
The graphs shown correspond to $\delta=-0.5$. The short wavelength boundary of the both
plots corresponds to the point where $\beta_0$ becomes imaginary.}
\label{h}
\end{figure}

We fix the detuning and  plot
$h$ and $\tilde h$  as  functions of the resonance wavelength, $\lambda_a$ in  Fig. \ref{h}.
On the short wavelength side the plots in Fig. \ref{h} are limited by the zero of the
denominator of $\beta_0$.
Through this it is seen that the two approaches
give qualitatively similar dependencies  in the long wavelength limit, whilst
in the short wavelength limit,
our calculations predict a significantly higher nonlinearity enhancement. Physically, one can
identify two factors determining changes in $h$ with the resonance wavelength.
Tendency for   ${\mathrm Re} h$ and $\tilde h$
to decrease with  decreasing $\lambda_a$ is linked to the fact
that  SPP intensity on the metal side increases
relative to the intensity on the dielectric side, as wavelength decreases.
Since the metal is linear in our model, it should lead to a drop in
the nonlinearity enhancement coefficient.
However the  smaller wavelengths reaching the SPP resonance
imply that the SPP field profile is getting squeezed closer to the interface on  both sides
and therefore the nonlinear part of the boundary conditions  becomes more important.
This  makes the difference between
${\mathrm Re}h$ and $\tilde h$, and the deviation of ${\mathrm Im}h$ from zero
pronounced in the short wavelength limit, see Fig. \ref{h}.

\section{Filamentation and transversely localized SPPs}
\subsection{Filamentation of SPPs}
The plane wave solution of Eq. (\ref{gl}) is
\begin{eqnarray}
&& A_0=\rho\times  \exp\left[i\frac{z}{2\beta_0}(\mathrm{Re}f
-\rho^2\mathrm{Re}\gamma)\right],\label{d0}\\ \nonumber && \rho^2=-\frac{\mathrm{Im}f}{\mathrm{Im}\gamma}>0.
\end{eqnarray}
$-\mathrm{Im}~f>0$ implies the gain above threshold, i.e.,
$\alpha>\alpha_0$. $\mathrm{Im}\gamma>0$ implies absorptive
nonlinearity compensating for the excess gain. Together these conditions
lead to the existence of the SPP solution with the
constant stationary amplitude, $\rho$. The expression for $\rho$, however,
diverges and our approach breaks down if the nonlinear absorption becomes zero, i.e.
$\mathrm{Im}\gamma=0$. This critical case requires one to
account for quintic nonlinear terms in the
perturbation expansion, which goes beyond our present objectives.

\begin{figure}
\centerline{\includegraphics[width=0.45\textwidth]{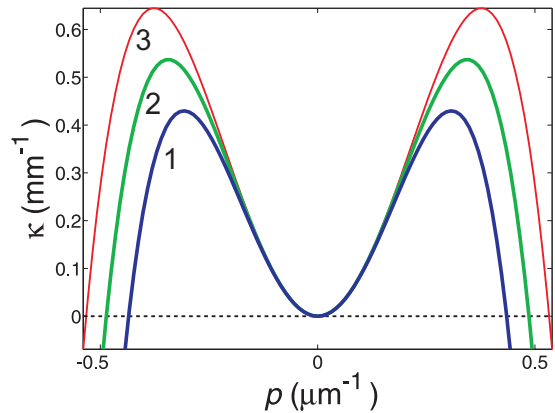}}
\caption{(Color online) Growth rate $\kappa$ of the filamentation
instability expressed in  physical units as a function of  momentum $p$.
$\lambda_a=594~nm$, $\delta=-0.3$, $\alpha_0=0.0063$ and $g=0.4,0.5,0.6$
for blue (line 1), green (line 2) and red (line 3) curves respectively.}
\label{Gain_Spectrum}
\end{figure}

\begin{figure}
\centerline{\includegraphics[width=0.45\textwidth]{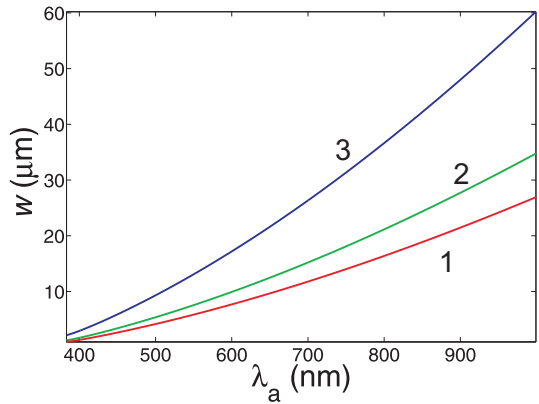}}
\caption{(Color online)
The characteristic filament size $w$ scaled back into 
physical units vs $\lambda_a$ for $\delta=-0.5$,
$g=0.1,0.3,0.5$ for blue (line 3), green (line 2) and red (line 1) curves respectively.}
\label{pmax}
\end{figure}

The solution (\ref{d0}) can  be unstable with respect
to a pattern forming filamentation instability, as  known
for the generic Ginzburg-Landau equation \cite{Aranson2002}.
In order to study the stability problem  we perturb $A_0$ with small amplitude
waves carrying transverse momentum $p$:
\begin{equation}
A=(1+q_+e^{\kappa z+ipy}+q_-^*e^{\kappa^* z-ipy})A_0\label{A1}.
\end{equation}
Inserting Eq.(\ref{A1}) into Eq.(\ref{gl}) and linearizing for small $|q_{\pm}|$
we find two solutions for $\kappa$. The unstable one is given by:
\begin{eqnarray}
&& 2\beta_0\kappa=\mathrm{Im}f+\sqrt{\mathrm{Im}^2f-p^2(p^2-2p^2_{max})}.
\end{eqnarray}
The filamentation instability sets in providing $\mathrm{Re}\kappa >0$.
In Fig. \ref{Gain_Spectrum}, it is seen that the  $\mathrm{Re}\kappa$ vs $p$ plot
has the typical two peak shape.
The maximal instability growth rate is achieved for
\begin{equation}
p=\pm p_{max},~p_{max}^2\equiv\rho\mathrm{Re}\gamma ~.
\end{equation}
The characteristic filament size in physical units is
$w\approx \lambda_{vac}/p_{max}$. $w$ as a function of $\lambda_a$ is plotted in Fig. \ref{pmax}.

\begin{figure}
\centerline{\includegraphics[width=0.45\textwidth]{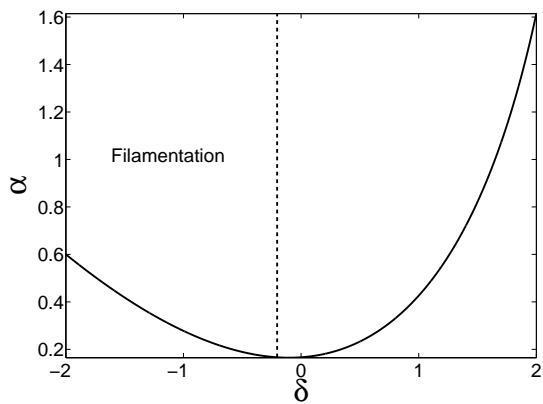}}
\caption{Lossless SPPs exist above the full line corresponding to $\alpha=\alpha_0$.
SPPs are unstable with respect to filamentation on the left from the dashed vertical line,
corresponding to $\mathrm{Re}\gamma=0$. $\lambda_a=400~nm$.}
\label{offset}
\end{figure}

The instability domain in the  ($\delta,\alpha$)-plane is shown  in
Fig. \ref{offset}. Filamentation is present for the self-focusing effective nonlinearity, i.e., $\mathrm{Re}\gamma>0$.
If the nonlinearity enhancement factor is  real,
then $\mathrm{Re}\gamma=0$ simply implies $\mathrm{Im}\chi=0$, which  is achieved
at the line center $\delta=0$. In this case,  the  nonlinearity  changes from
 focusing (filamentation) to defocusing (no filamentation)
at the  atomic resonance, i.e. exactly as  in the bulk material.
However, the fact that $\mathrm{Im}h\ne 0$ leads to the  
offset of the instability boundary away from $\delta=0$, see Fig. \ref{offset}.
Gain values corresponding to approximately $50$\% above  threshold
imply the development of a filamentary pattern over  distances of $1-3$mm.

\subsection{Bright and dark localized SPPs}
The cubic Ginzburg-Landau equation is known to have a wide variety of
localised  solutions, which
can be relevant in the SPP context under  different circumstances.
Detailed classification of these solutions can be found in, e.g., Refs. \cite{Aranson2002,Akhmediev1997}.
Here we briefly introduce the most ubiquitous of those, which are
bright Pereira-Stenflo \cite{Pereira1977} and dark Nozaki-Bekki \cite{Nozaki1984}
localised solutions.

Both bright and dark localised solutions exist  under the same
conditions: $\mathrm{Im}f<0$ (positive gain) and $\mathrm{Im}\gamma
>0$ (nonlinear absorption). The  bright solution  is given by
\begin{equation}
A(y,z) = \rho\sqrt{3\over 2}\left[sech(Ky)\right]^{1+ia}exp(iu z),
\label{bright}\end{equation}
and  the dark one by
\begin{equation}
A(y,z) = \rho\frac{tanh(sy)}{\left[cosh(sy)\right]^{ib}}exp(ivz).
\label{dark}\end{equation}
Explicit expressions for the parameters entering Eqs. (\ref{bright}), (\ref{dark})
are given in the Appendix.
In the limit $|y|\to\infty$ the dark solution tends towards the plane wave solution (\ref{d0}).

\begin{figure}
\centerline{\includegraphics[width=0.45\textwidth]{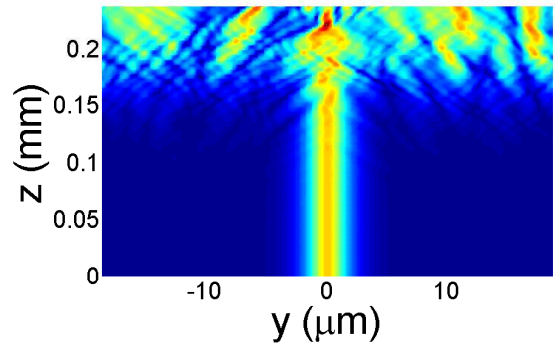}}
\caption{(Color online) Propagation of a bright localized solution and
 instability of its background: $\lambda_a=594~nm$, $\delta=-0.3$, $\alpha=0.0183$, $\alpha_0=0.0063$.}
\label{br}
\end{figure}
\begin{figure}
\centerline{\includegraphics[width=0.45\textwidth]{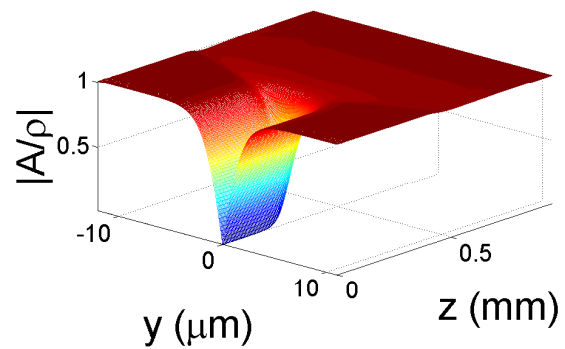}}
\caption{(Color online) 
Destabilization of the dark soliton due to core instability: 
$\lambda_a=594~nm$, $\delta=-0.3$, $\alpha=0.0183$, $\alpha_0=0.0063$.}
\label{da}
\end{figure}

The bright solution is unstable because its zero background is
unstable above  threshold. This instability is relatively slow
to develop and practical observation of bright solitons over
distances of $100$s of microns is still possible, see numerical
modelling results in Fig.  \ref{br}. The dark solution is known to
be unstable with respect to the core instability through  most of
its existence domain, see, e.g., \cite{Chate1992}, which is
complemented by the filamentation of  background, provided the
effective Kerr nonlinearity is self-focusing. Fig. \ref{da} shows
an example of the core instability, where one can see that for the chosen
parameters it develops over the shorter distance, if compared to the
instability of the zero background of the bright solution in Fig.  \ref{br}.

\section{Summary}
We have considered nonlinear propagation of the amplified and
diffracting surface plasmon polaritons above  the  threshold beyond
which the plasmon propagation constant becomes real.  Starting from
the first principle Maxwell equations, we have developed a technique
allowing derivation of the  complex cubic Ginzburg-Landau equation
for the slowly varying plasmon amplitude. The nonlinear plasmon
solutions found by our method  is guaranteed to satisfy nonlinear
boundary conditions at the interface to the required accuracy. This
distinguishes our approach and results from the recently proposed
derivation of the nonlinear Schr\"odinger equation for surface
plasmons, which satisfy  boundary conditions only in the linear
approximation \cite{Feigenbaum2007,Davoyan2009}. We have found that
the   nonlinearity enhancement factor is always complex and hence 
mixes real and imaginary parts of the  intrinsic nonlinearity of a
dielectric. This  mixing changes conditions required for the filamentation
of plasmons and  the existence of the dark (Nozaki-Bekki) and
bright (Pereira-Stenflo) spatially localized waves, relative to the
respective conditions in  bulk medium. Though both of the localized 
solutions are unstable with respect to growth of small perturbations,
the bright ones demonstrate quasi-stable propagation over the  distances
of $100$'s of microns and  are likely to be practically observable.
Finding mechanisms leading to stabilization of 
these structures is an important topic for future research.   

\section{Acknowledgement}
We acknowledge useful discussions  with A. Gorbach, A. Zayats.

\section*{Appendix}
The parameters entering Eq. (\ref{bright})
are expressed in terms of the parameters for the Ginzburg-Landau equation Eq. (\ref{gl}) as
\begin{eqnarray}
\nonumber a & = & -\frac{3\mathrm{Re}\gamma}{2\mathrm{Im}\gamma}+\sqrt{2+\left(\frac{3\mathrm{Re}\gamma}
{2\mathrm{Im}\gamma}\right)^2}, \\
\nonumber  K^2 & = & -\frac{1}{2a}\mathrm{Im}f, \\
\nonumber  u & = & \frac{1}{2\beta_0}\mathrm{Re}f+\frac{a^2-1}{4\beta_0a}\mathrm{Im}f.
\end{eqnarray}

Parameters entering Eq. (\ref{dark}) are
\begin{eqnarray}
\nonumber  b & = & -\frac{3\mathrm{Re}\gamma}{2\mathrm{Im}\gamma}-\sqrt{2+\left(\frac{3\mathrm{Re}\gamma}{2\mathrm{Im}\gamma}\right)^2},  \\
\nonumber  s^2 & = & \frac{1}{3b}\mathrm{Im}f, \\
\nonumber  v & = & \frac{1}{2\beta_0}\mathrm{Re}f-\frac{1}{3b\beta_0}\mathrm{Im}f.
\end{eqnarray}


\end{document}